\documentclass[twocolumn,showpacs,preprintnumbers,amsmath,amssymb]{revtex4}

\usepackage[dvips]{graphicx}
\usepackage{dcolumn}
\usepackage{bm}

\begin{document}


\title{Strong tunable coupling between a superconducting charge and phase qubit}

\author{A. Fay$^1$, E. Hoskinson$^1$, F. Lecocq$^1$, L. P. L\'evy$^1$, F. W. J. Hekking$^2$, W. Guichard$^1$ and O. Buisson$^1$}

\affiliation{$^1$Institut N\'eel, C.N.R.S.- Universit\'e
Joseph Fourier, BP 166, 38042 Grenoble-cedex 9, France}

\affiliation{$^2$LPMMC, C.N.R.S.- Universit\'e
Joseph Fourier, BP 166, 38042 Grenoble-cedex 9, France}

\date{\today}

\begin{abstract}

We have realized a tunable coupling over a large frequency range
between an asymmetric Cooper pair transistor (charge qubit) and a
dc SQUID (phase qubit). Our circuit enables the independent
manipulation of the quantum states of each qubit as well as their
entanglement. The measurements of the charge qubit's quantum
states is performed by resonant read-out via the measurement of
the quantum states of the SQUID. The measured coupling strength is
in agreement with an analytic theory including  a capacitive and a
tunable Josephson coupling between the two qubits.

\end{abstract}

\pacs{Valid PACS appear here}
\maketitle

\begin{figure}
\includegraphics[scale=0.4]{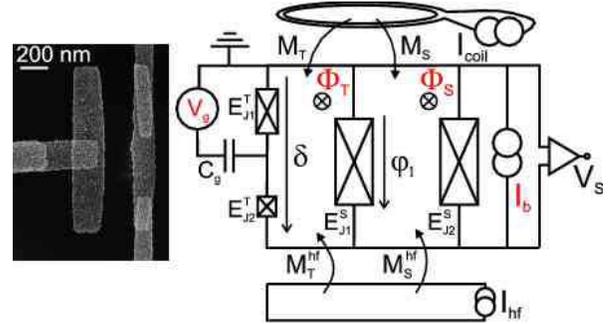}
\caption{ Electrical schematic of the coupled circuit. The working
point is controlled by a dc gate voltage $V_g$, a current biased
$I_b$ and the fluxes $\Phi_T$ and $\Phi_S$. The current $I_{coil}$
and $I_{hf}$ produces a flux on the SQUID and transistor through
respectively the mutual $M_S$, $M_T$ and $M_S^{hf}$, $M_T^{hf}$.
The high frequency (hf) line is also used to produce a $\mu$w flux
pulse and a nanosecond flux pulse for the escape measurement of
the SQUID. On the left side a SEM image of the asymmetry Cooper
pair transistor.}. \label{Proceduremesure}
\end{figure}

Interaction between two quantum systems induces entangled states
whose properties have been studied since the 80's for pairs of
photons \cite{Dalibard_PRL82}, for atoms coupled to photons
\cite{Raimond_RMP01} and for trapped interacting
ions\cite{Leibfried_RMP03}. In the last decade, quantum
experiments were extended to macroscopic solid state devices
opening the road for application within the field of quantum
information. In superconducting circuits, theoretical proposals
\cite{Buisson_00, Plastina_PRB03, Blais_PRA04} and experimental
realizations on interacting quantum systems were put forward. In
these systems coupling has been achieved between a quantum
two-level system (qubit) and a
resonator\cite{Chiorescu_Nature04,Wallraff_Nature04,Johansson_PRL06}
as well as between two identical
qubits\cite{Pashkin_Nature03,Berkley_Science03,McDermott_Science05}.
In these pioneering circuits the interaction between the quantum
systems was realized through a fixed capacitive or inductive
coupling. The tunability of the coupling strength appears as an
important issue to optimize the control of two or more coupled
quantum systems. Indeed it enables to decouple the quantum systems
for individual manipulations and to couple them when entanglement
between the quantum states is needed. Recently different tunable
couplings between two identical qubits have been proposed and
measured\cite{Averin_PRL03,Niskanen_Science07,Hime_Science06,Sillanpaa_Science07,Majer_Nature07}.
In this Letter we report for the first time on a tunable composite
coupling between a charge qubit, an asymmetric Cooper pair
transistor (ACPT) and a phase qubit, a dc SQUID. In our circuit
(see Fig.\ref{Proceduremesure}) the coupling is composed of two
independent terms, a fixed capacitive and a tunable Josephson
part, leading to a tunability of the total coupling.


The dynamics of the current biased dc SQUID can be described by
the Hamiltonian of an anharmonic oscillator:
$\hat{H}_{S}=\frac{1}{2}h \nu _{p} (\hat{P}^2+\hat{X}^2)- \sigma h
\nu_{\rm p}\hat{X}^3 \label{Hamiltonian}$ where $\nu_p$ is the
plasma frequency of the SQUID. Here $\hat{P}$ and $\hat{X}$ are
the reduced charge and phase conjugate
operators\cite{Claudon_PRL04}. In our case the anharmonicity
prevents multi-plasmon excitation and therefore the system at low
energy reduces to a two-level system with levels denoted by $|0
\rangle$ and $|1 \rangle$ corresponding respectively to the zero-
and one-plasmon state. At low energies the SQUID Hamiltonian
therefore reads $\hat{H}_{S}=h \nu _{S}\hat{\sigma}_z^ {S}\,/2$
where $\hat{\sigma_z}$ is the Pauli matrix. The frequency between
these two levels $\nu_S$ depends on the working point and is
determined by the dc flux $\Phi_S$ through the SQUID loop and the
bias current $I_b$. The ACPT can be described as a two-level
system with quantum states denoted by $|-\rangle$ and $|+\rangle$
for respectively the ground and first excited state. The
hamiltonian of the ACPT takes the form $\hat{H}_{T} = h \nu_{T}
\hat{\sigma}_z^ {T}\,/2$ where $\nu_{T}$ depends on the
gate-induced charge $n_g$ and the phase difference $\delta$ across
the transistor. We now turn to the coupling of both quantum
systems in a circuit shown in Fig.\ref{Proceduremesure}. As the
ACPT is in parallel to the SQUID, both a Josephson and a
capacitive coupling appear between these two quantum systems. The
Josephson coupling results from the phase relation along the loop
between the transistor bias and the closer SQUID junction. The
capacitive coupling is explained by the charge displacement
between the transistor and SQUID capacitance.
 The total coupling can be tuned in our circuit from about $1.2GHz$ down to $0.1GHz$.

The ACPT consists of a superconducting island connected by two
Josephson junctions of different surfaces of about $0.02\
\mu$m$^2$ and $0.05\ \mu$m$^2$, respectively to the
supercondcuting electrodes. The dc SQUID comprises two large
Josephson junctions of $5\ \mu$m$^2$ area each, enclosing a $347\
\mu$m$^2$ superconducting loop. The ACTP and the SQUID Josephson
junction closer to the ACPT realizes a second loop of $126\
\mu$m$^2$ surface.  The coupled circuit is realized by a three
angle shadow evaporation of aluminum with two different oxydations
respectively for the SQUID junctions and the ACPT junctions.
Measurements are performed in a dilution fridge at $T=30\:
\text{mK}$. The microwave ($\mu$w) flux and charge-gate signal are
guided by $50\: \Omega$ coax lines and $40dB$ attenuated at low
temperature before reaching the circuit through a mutual
inductance and the gate capacitance, respectively. The measurement
of the quantum states of the circuits is performed by a nanosecond
flux pulse which produces switching to the voltage state of the
SQUID\cite{Claudon_PRB07}.

\begin{figure}[htbp!]
  \centering
  \includegraphics[width=0.8\linewidth]{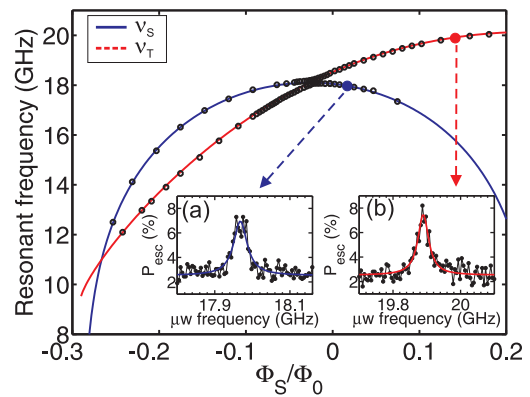}
  \caption{ Experimental resonant frequency versus $\Phi_S$ for the coupled circuit  with $I_b=1890$ nA and $n_g=1/2$. The blue and red solid lines are the fit using the uncoupled  hamiltonians of  the dc SQUID and the ACPT respectively. Inset: Escape probability of the SQUID versus frequency probing (a) the SQUID at $\Phi_S=0.02\ \Phi_0$  and (b) the ACPT at $\Phi_S=0.14\ \Phi_0$ (fitted by a lorentzian law).}
 \label{energie}
\end{figure}

We first study the individual resonant frequency of the SQUID and
the ACPT (Fig.~\ref{energie}).
  Spectroscopy measurements of the SQUID are performed by a $\mu$w
flux pulse followed by a nanosecond flux pulse. The escape
probability shows a resonant peak associated with the transition
$|0\rangle \rightarrow |1\rangle$ (inset (a) of
Fig.~\ref{energie}). The SQUID resonance frequency $\nu_{S}$ can
be tuned from 8GHz to more than 20GHz as a function of the bias
current $I_b$ and the magnetic flux $\Phi_S$ in the SQUID loop.
>From flux calibration, we obtain $M_S=0.13$ pH and $M_S^{hf}=1.58$
pH. From the measured resonance frequency $\nu_{S}$ the SQUID
parameters such as $E_J^{S}$, $E_C^{S}$
 and the total SQUID inductance $L$ can be determined
with a precision better than $1 \% $. We find a critical current
of $I^{S}_{c}=1346nA$, a capacitance $C^{S}=0.227pF$ per junction,
an inductance $L=190pH$ and an inductance asymmetry of $\eta=0.29
$ between the two SQUID arms. These values are similar to typical
parameters of previous samples\cite{Claudon_PRB07}. When the
SQUID's working point frequency increases from 8GHz to 20GHz the
resonance width changes from 200MHz to 20MHz. The finite width is
consistent with a 10nA RMS current noise and a $1m\Phi_0$ RMS flux
noise\cite{Claudon_PRB06}.
    Rabi-like oscillations have been measured with a typical
decay time of about 10ns and a relaxation time of about 30ns.
These times are shorter in comparison to our previous SQUID
sample. Moreover a high density of parasitic resonances is
observed in the current sample (see Fig. 9 of
Ref.~\cite{Claudon_PRB07}) which could explain these shorter
times. The origin of these resonances is still not completely
understood but has been already observed in other phase-qubits
\cite{Cooper_PRL04}. All presented measurements have been done
at working points where these parasitic resonances are not visible
through spectroscopy measurements.

The energy levels of the ACPT can be determined as well by escape
probability measurements on the SQUID via resonant read out. We
apply a $\mu$w signal of 1$\mu$s on the gate line at fixed
frequency when the ACPT and the SQUID are off resonance. If the
applied $\mu$w frequency matches the ACPT frequency the $|+
\rangle$ level of the ACPT is populated. For the measurement a
nanosecond flux pulse with a rise time of 2 ns drives the two
systems adiabatically across the resonance where the coupling is
about 1 GHz (see below). The initial state $|+,0 \rangle$ is
thereby transferred into the state $|-,1 \rangle$
\cite{Buisson_PRL03}. Afterwards an escape measurement is
performed on the SQUID (Inset b of Fig.\ref{energie}).

\begin{figure}[htbp!]
  \centering
  \includegraphics[width=0.8\linewidth]{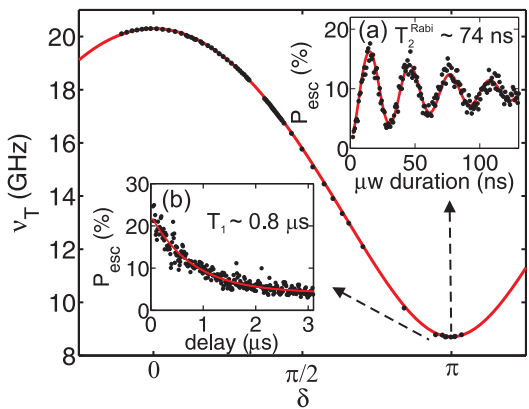}
  \caption{The ACPT energy versus $\delta$ at $n_g=0.5$ fitted by the ACPT hamiltonian.
Inserts: Measurements at $\delta=\pi$. (a) Escape probability
versus $\mu$w pulse duration for $-3dBm$ room
  temperature $\mu$w power. (b) Escape
  probability versus delay time between the $\mu$w and the measurement pulse fitted
  by an exponential decay (continuous line) giving $T_1=810ns$.}
  \label{ACPT}
\end{figure}

 The ACPT resonant frequency as a function of $\delta$ at $n_g=1/2$ is shown
in Fig.~\ref{ACPT}. Here $\delta$ is given by the relation
 $\delta=\varphi_1+L_1\,I_c^{S}sin(\varphi_1)/\Phi_0-2\pi \Phi_T/\Phi_0$ where is $\Phi_T$ the dc flux inside the loop,
 $\phi_1$ the phase
 difference across the SQUID junction closer to the transistor and $L_1$ the inductance of the corresponding branch of the SQUID.
 In our set-up we have $M_T=0.047$ pH and $M_T^{hf}=0.35$ pH and $L_1=70$
 pH.
 The qubit resonant frequency $\nu_{T}$ versus
$\delta$ can be fitted within $1 \% $ error by considering that
the $|+ \rangle$ and $|- \rangle$ states are superpositions of
four charge states. The ACPT has two optimal working points for
qubit manipulations. The one at ($n_g,\delta$)=(1/2,0) was
extensively studied in the Quantronium symmetric
transistor\cite{Vion_Science02}. The ($n_g,\delta$)=(1/2,$\pi$)
working point appears as a new optimal point created by the
asymmetry of the transistor. The width of the resonance peak far
from the optimal points is typically $40$ MHz while close to the
two optimal points $\delta=0$ and $\delta=\pi$, it is typically
around 20MHz.
    From the two extreme resonant frequencies $\nu_{T}=20.302$ GHz and  $\nu_{T}=8.745$ GHz, the critical
current of the two junctions can be deduced and we obtain
$I^{T}_{c,1}=30.1nA$ and $I^{T}_{c,2}=12.3nA$. From the frequency
spectrum $\nu_{T}$ versus the gate charge $n_g$, we find a total
transistor capacitance of $C^{T}=2.9fF$ and a gate capacitance
$C_{g}=29aF$. Fig.\ref{ACPT}a presents Rabi oscillations in the
ACPT at the new optimal point ($n_g,\delta$)=(1/2,$\pi$). The Rabi
frequency follows a linear dependance on the $\mu$w amplitude
 as expected for a two-level quantum system. The
two level system presents a long relaxation time of about $800ns$
(Fig.~\ref{ACPT}b).

\begin{figure}[htbp!]
  \centering
 \includegraphics[width=0.85\linewidth]{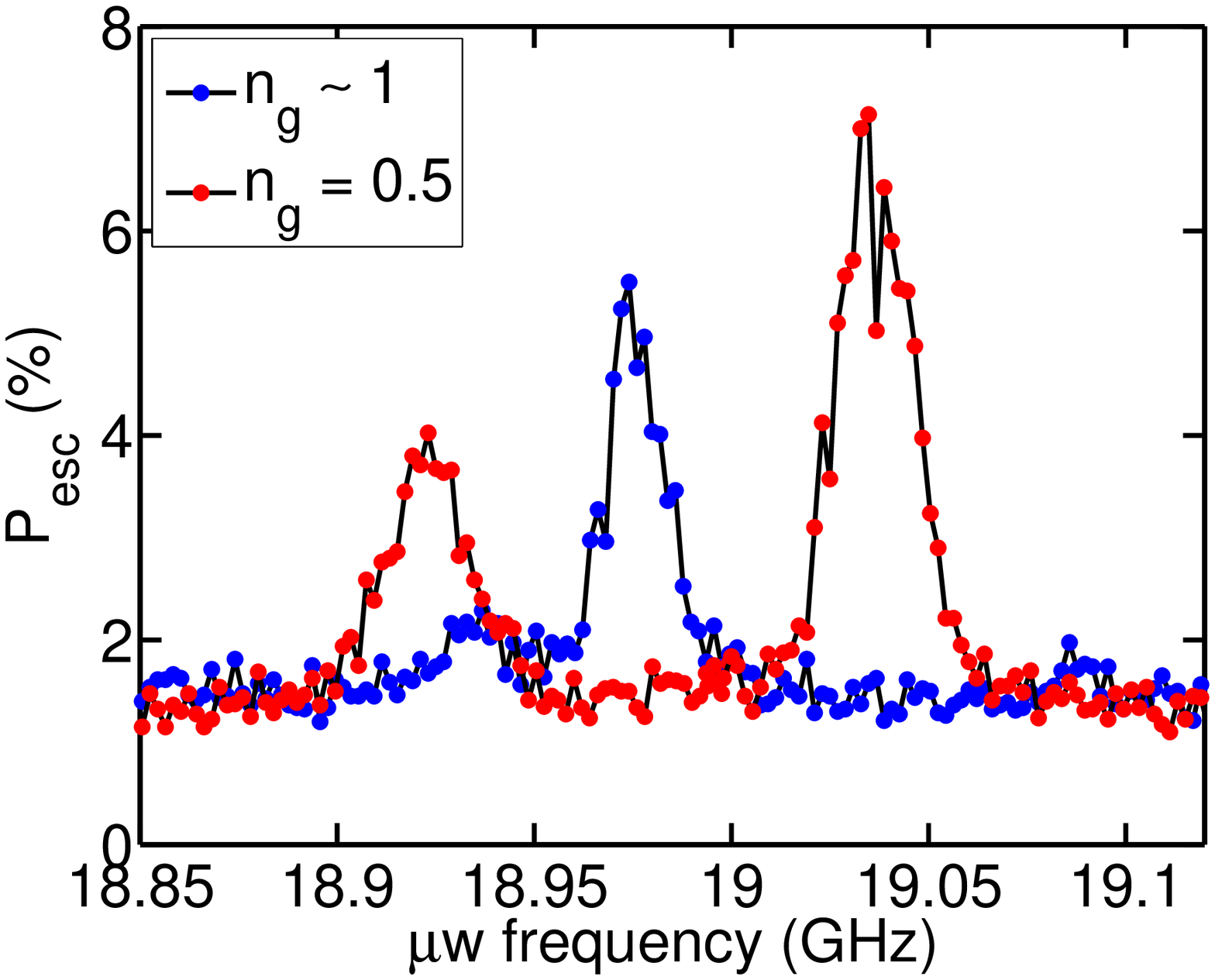}
  \includegraphics[width=0.85\linewidth]{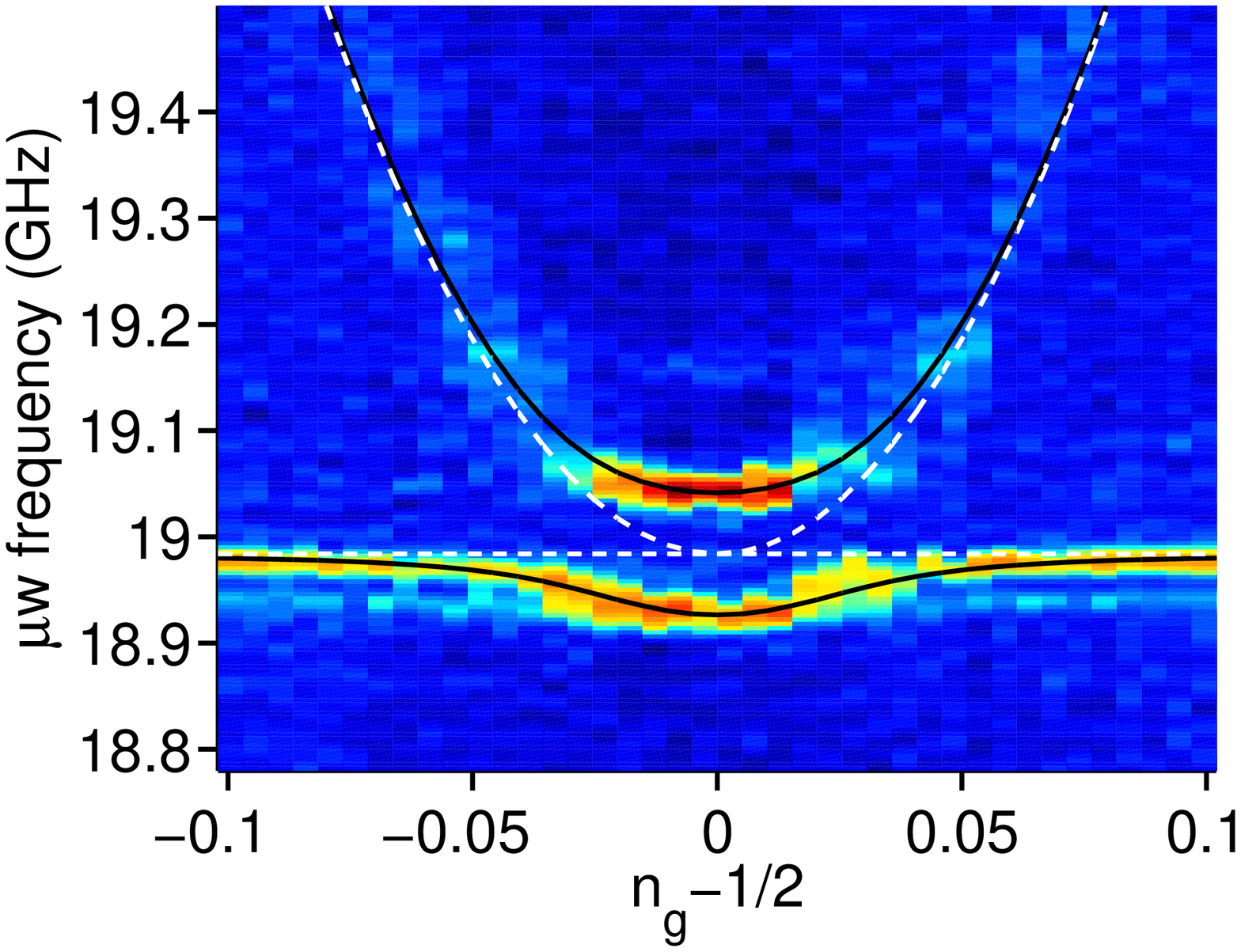}
 \includegraphics[width=0.85\linewidth]{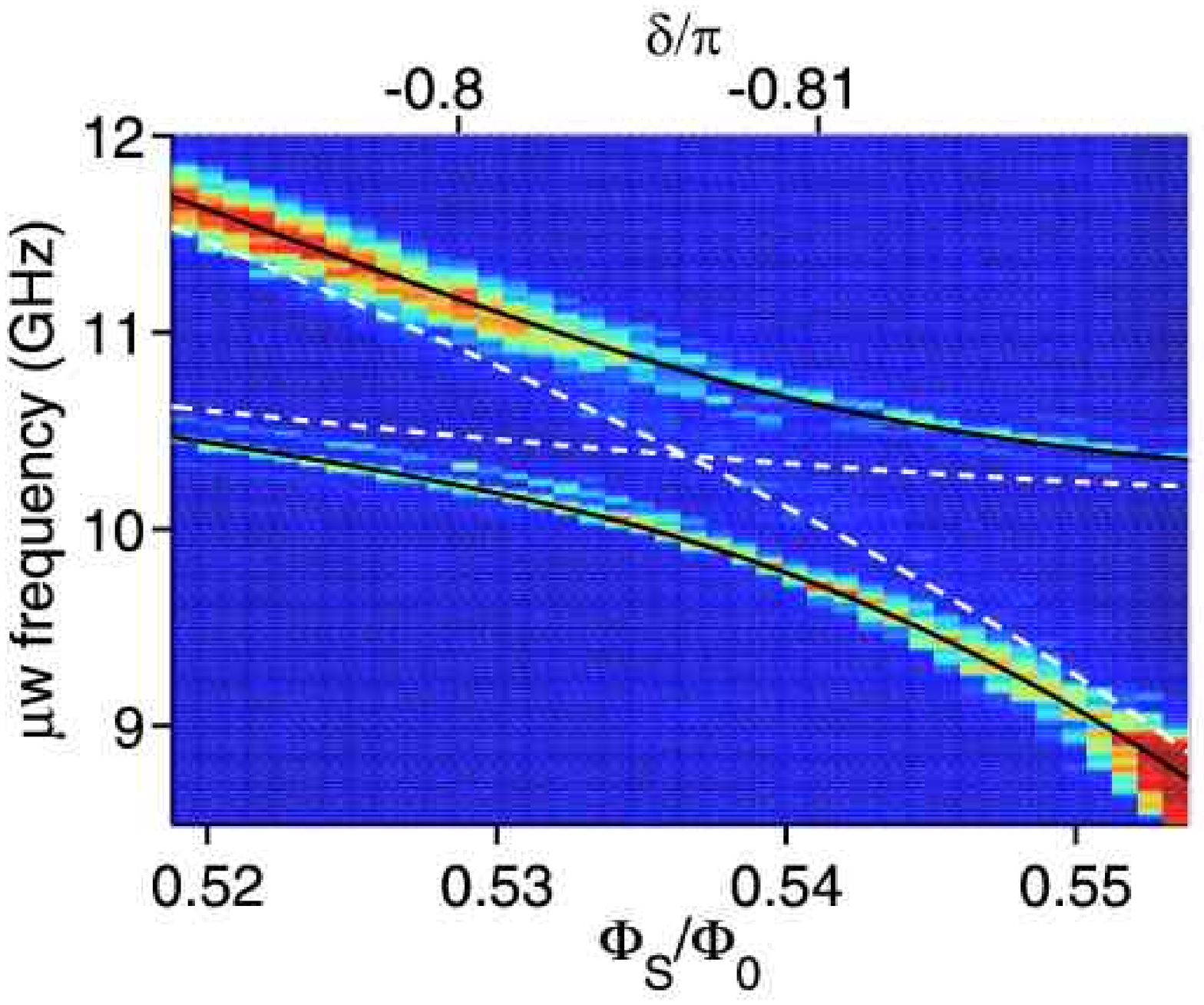}
 \caption{(a)Escape probability of the SQUID at the working point
  $I_b= 1647$ nA, $\Phi_S=0.03\ \Phi_0$, $\delta=0.26\pi$. Blue curve: At $n_g\sim1$ SQUID and ACPT are out off
  resonance ($\nu_T\sim31.6$ GHz). Red curve: At $n_g=0.5$ the resonance condition is
  fulfilled leading to antilevel-crossing.(b) $P_{esc}$ versus $n_g$ and $\mu$w frequency at the
 working point $I_b= 1647$ nA, $\Phi_S=0.03\ \Phi_0$ and $\delta=0.26\pi$.
 ( c) $P_{esc}$ versus $\Phi_S$($\delta$) and $\mu$w frequency at the working point
 $I_b= 107$ nA and $n_g=1/2$. Blue color corresponds to small $P_{esc}$ , red color to large $P_{esc}$.
 Dashed and continuous lines correspond to uncoupled and coupled cases.}
  \label{2Dplot}
\end{figure}

Hereafter we consider the case when the two qubits are in
resonance ($\nu_{T}= \nu_{S}$). Fig.~\ref{2Dplot}a shows the
measured escape probability at the working point $I_b= 1647$ nA
and $\Phi_S=0.03\ \Phi_0$  for two different gate charges
 $n_g=1/2$ and $n_g \sim 1$ corresponding respectively to the in and off resonance case.
 Off resonance, the ACPT frequency being very much larger than the SQUID resonance, only one
 resonance peak is observed which corresponds to
the $|1\rangle$ state excitation of the SQUID. At $n_g=1/2$  the
resonance condition between the ACPT and the SQUID is satisfied
for this working point. The coupling between the two systems leads
to a splitting of the resonance peak of about $120$ MHz into two
peaks corresponding to the two entangled states $|0,+\rangle \pm
|1,-\rangle$. The resonance width is about four times thinner than
the coupling strength which demonstrates clearly the strong
coupling of the ACPT two-level system with the zero- and the
one-plasmon state of the dc SQUID. In Fig.~\ref{2Dplot}b, the
escape probability versus $n_g$ and $\mu$w frequency is plotted at
the same working point. Far from the resonance condition the
 value can be well estimated assuming two uncoupled circuits. In
 the vicinity of $n_g=1/2$, anti-level crossing occurs
 modifying the individual resonance frequency of the two circuits.
 In Fig.~\ref{2Dplot}c, the escape probability versus $\Phi_S$ and $\mu$w frequency is measured at $n_g=1/2$ at a different working point.
 Anti-level crossing is clearly observed with a splitting of about $900$MHz.
 The width of the two resonances strongly depends on $\Phi_S$ and varies from $200$MHz to about $40$MHz as
 the crossing point is passed.
 This effect can be explained by the large difference of the resonance width of the SQUID and the ACPT around this working point.

\begin{figure}[htbp!]
  \centering
  \includegraphics[width=0.8\linewidth]{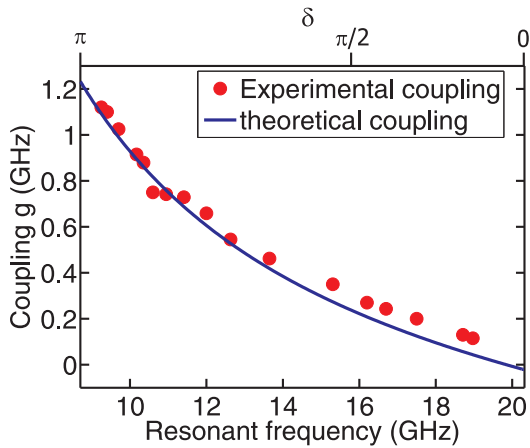}
  \caption{Coupling strength versus the frequency at the resonance
  condition between the two circuits at $n_g=1/2$. The points are
  experimental data and the continuous line is the theoretical prediction.}
  \label{Coupling}
\end{figure}

The coupling strength between the two qubits is measured at $n_g=1/2$ and at the working points where
 the resonance condition $\nu_{T}= \nu_{S}$ is satisfied.
The frequency splitting is plotted versus the resonant frequency in Fig.\ref{Coupling}.
 The coupling is minimal at
$\nu_{T}=20.3GHz$ and strongly increases with decreasing resonant
frequency up to a maximum value of 1.2 GHz. Note that when the
resonance frequency changes from 20.3 GHz down to 8.8 GHz the
phase bias over the ACPT changes from $\delta=0$ to $\delta=\pi$.
We find therefore nearly zero coupling at $\delta=0$ and a very
strong coupling of 1.2 GHz at $\delta=\pi$.

For the theoretical analysis we consider that the transistor
$|-\rangle$ and $|+ \rangle$ states are superpositions of two
charge states and we neglect anharmonicity effects of the SQUID
potential on the frequency $\nu_{S}$.
We obtain the following analytical expression for the coupling
strength at a gate charge of $n_g=1/2$:
$H_{coupling}=(E_{c,c}/4)\sigma_S^{x}\sigma_T^{x}-E_{c,j}/2(\cos(\chi-\delta/2)\sigma_S^{y}\sigma_T^{y}+\sin(\chi-\delta/2)\sigma_S^{y}\sigma_T^{z}/2)$,
where $E_{c,c}= (1-\lambda)\sqrt{E_{C}^{S}/h \nu_p} h \nu_p$ and
  $E_{c,j}= (1-\mu)\sqrt{E_{C}^{S}/h \nu_p}E^{T}_{j} $
  with $\lambda=(C^{T}_{1}-C^{T}_{2})/(C^{T}_{1}+C^{T}_{2})$
  and $\mu=(E^{T}_{J,1}-E^{T}_{J,2})/(E^{T}_{J,1}+E^{T}_{J,2}$) being the transistor capacitance
  and Josephson energy asymmetry, respectively. $E_{C}^{S}\approx e^2/2C^{S}$ with
  $C^{S}$ the SQUID capacitance, $E^{T}_{j}=E^{T}_{J,1}+E^{T}_{J,2}$ the transistor Josephson energy and $\tan(\chi)=-\mu\, \tan(\delta/2)$.
   The coupling contains two independent contributions: one related to the capacitance and the other one
 to the Josephson coupling of the ACPT.
Close to resonance, slow dynamics dominates and the hamiltonian
simplifies to a Jaynes-Cummings type Hamiltonian $H_{coupling}=
\frac{1}{2}g(\sigma_S^{+}\sigma_T^{-} + \sigma_S^{-}\sigma_T^{+})$
where $g=(E_{c,c}/2-E_{c,j}\cos(\chi-\delta/2))$ and
$\sigma_{S/T}^{+/-}$ creates or annihilates an excitation in the
SQUID or the ACPT. At this point we stress that the coupling
strength at $n_g=1/2$ depends only on the $\delta$ parameter. If
we replace one of the transistor junctions by a pure capacitance
($E^{T}_{J,2}=0$) we obtain $E_{c,j} =0$ and we retrieve the
capacitive coupling \cite{Buisson_00} calculated for a Cooper pair
box coupled to a SQUID. For a \emph{symmetric} transistor
($\lambda=\mu=0$) the charge and the Josephson coupling compensate
each other, giving zero coupling for any value of the $\delta$
parameter. It is the asymmetry of the transistor which enables non
zero coupling at the optimum point of the charge qubit. In
particular, for the case that $\lambda=\mu$ - which is realized
for a transistor containing two junctions having the same plasma
frequency - the total coupling vanishes at $\delta=0$ but becomes
non zero at the second optimum point at $\delta=\pi$. By assuming
an asymmetry of $\mu=\lambda=41.9\%$ for our sample the coupling
strength can be very well fitted without any other free parameters
as can be seen in Fig. \ref{Coupling}. The slight discrepancy can
be explained by a small difference between $\lambda$ and $\mu$.

    In conclusion, we have demonstrated strong tunable coupling between two
superconducting qubits. Far from resonance our quantum circuit
enables us to control the quantum dynamics of each qubit
separately. At resonance we demonstrate entanglement between the
quantum states of the charge and phase qubit which is consistent
with the exchange of a single energy quantum. The measured
coupling strength could be perfectly understood by an analytical
coupling expression of the type Jaynes-Cumings Hamiltonian. The
quantum state measurement of the charge-phase qubit has been
performed via resonant readout by measuring the quantum state of
the SQUID. Our result encourages the future development of quantum
information processing in solid-state devices.

We thank  for fruitful discussions. This work was supported by two
ACI programs, by the EuroSQIP project and by the Institut de
Physique de la Mati\`ere Condens\'ee.

\end{document}